\documentclass[prl,nofootinbib,superscriptaddress,twocolumn]{revtex4-1}
\usepackage[a4paper,left=1.5cm,right=1.5cm,top=3cm,bottom=3cm]{geometry}

\usepackage{verbatim}
\usepackage{color}
\usepackage{amsfonts,amssymb,mathrsfs,amsmath}
\usepackage{framed}
\usepackage{mdframed}
\usepackage{latexsym}
\usepackage{graphicx}
\usepackage{datetime}
\newdateformat{mydate}{\THEDAY{ }\monthname[\THEMONTH]{ }\THEYEAR}

\allowdisplaybreaks

\usepackage{tikz}
\usepackage{color}
\usepackage{framed}
\usepackage{hyperref}
\hypersetup{colorlinks, citecolor=bluscuro, linkcolor=black, urlcolor=bluscuro}
\definecolor{rossos}{cmyk}{0,1,1,0.55}
\definecolor{bluscuro}{rgb}{0.15, 0.2, .85}
\definecolor{bluchiaro}{cmyk}{1,.3,0.,0.1}

\newcommand{\be}{\begin{equation}}
\newcommand{\ee}{\end{equation}}
\renewcommand{\d}{{\rm d}}
\def\ii{{\text{\tiny i}}}
\def\PBH{\text{\tiny PBH}}
\newcommand{\llp}{\left [}
\newcommand{\rrp}{\right ]}
\newcommand{\lp}{\left (}
\newcommand{\rp}{\right )}

\def\lsim{\mathrel{\rlap{\lower4pt\hbox{\hskip0.5pt$\sim$}}
    \raise1pt\hbox{$<$}}}         
\def\gsim{\mathrel{\rlap{\lower4pt\hbox{\hskip0.5pt$\sim$}}
    \raise1pt\hbox{$>$}}}         

\newcommand{\arXiv}[2]{\href{http://arxiv.org/pdf/#1}{{\tt [#2/#1]}}}
\newcommand{\arXivold}[1]{\href{http://arxiv.org/pdf/#1}{{\tt [#1]}}}

\newcommand{\name}{GW190521}

\begin{document}

\title{The \name \ Mass Gap Event and the  Primordial Black Hole Scenario}

\author{V. De Luca}
\address{D\'epartement de Physique Th\'eorique and Centre for Astroparticle Physics (CAP), Universit\'e de Gen\`eve, 24 quai E. Ansermet, CH-1211 Geneva, Switzerland}

\author{V. Desjacques}
\address{Physics department and Asher Space Science Institute, Technion, Haifa 3200003, Israel}

\author{G. Franciolini}
\address{D\'epartement de Physique Th\'eorique and Centre for Astroparticle Physics (CAP), Universit\'e de Gen\`eve, 24 quai E. Ansermet, CH-1211 Geneva, Switzerland}

\author{P. Pani}
\address{Dipartimento di Fisica, 
Sapienza” Università 
di Roma, Piazzale Aldo Moro 5, 00185, Roma, Italy}
\address{INFN, Sezione di Roma, Piazzale Aldo Moro 2, 00185, Roma, Italy}

\author{A.~Riotto}
\address{D\'epartement de Physique Th\'eorique and Centre for Astroparticle Physics (CAP), Universit\'e de Gen\`eve, 24 quai E. Ansermet, CH-1211 Geneva, Switzerland}

\address{INFN, Sezione di Roma, Piazzale Aldo Moro 2, 00185, Roma, Italy}

\date{\today}

\begin{abstract}
\noindent
The LIGO/Virgo Collaboration has recently observed \name, the first binary black hole merger with at least 
the primary component mass in the mass gap predicted by the pair-instability supernova theory. This observation 
disfavors the standard stellar-origin formation scenario for the heavier black hole, motivating alternative hypotheses.
We show that \name\ cannot be explained within the Primordial Black Hole~(PBH) scenario if 
PBHs do not accrete during their cosmological evolution, since this would require an abundance which is already in 
tension with current constraints.
On the other hand, \name\ may have a primordial origin if PBHs accrete efficiently before the reionization epoch.
\end{abstract}

\maketitle

\paragraph{Introduction.}
\noindent
 The observation of gravitational waves~(GWs) from black-hole~(BH) mergers detected by the LIGO/Virgo 
Collaboration during the first observation runs (O1-O2-O3)~\cite{LIGOScientific:2018mvr,LIGOScientific:2020stg, Abbott:2020khf}
have resumed the interest in understanding the origin 
of the merging BH population~\cite{Barack:2018yly}. A fascinating possibility is represented by PBHs, whose formation 
took 
place in the early stages of the universe (see Ref.~\cite{sasaki}  for a review), and 
that can comprise a significant fraction $f_\PBH$ of the dark matter~(DM)~\cite{Sasaki:2016jop,Bird:2016dcv}. 
The possible primordial origin of at least some of the GW events has motivated several studies on the 
confrontation between the PBH scenario and the GW data \cite{Sasaki:2016jop,Bird:2016dcv,Clesse:2016vqa, 
Ali-Haimoud:2017rtz, raidal, raidalsm,ver, Gow:2019pok, Clesse:2020ghq,  Hall:2020daa}.

The recent release of \name\ \cite{newevents}, describing the coalescence of two BHs with masses $M_1 = 
85^{+21}_{-14} M_\odot$ and $M_2 =  66^{+17}_{-18} M_\odot$,  has assigned  to PBHs even more relevance due to the 
so-called  mass gap expected between about $65 M_\odot$ and $135 M_\odot$ in the spectrum of population I/II 
stellar-origin BHs. 
Pulsational pair instability (originated from unstable stellar cores due to the 
production of electron-positron pairs at high temperature) 
makes stars with a helium core mass in the range $\sim (32 \div 64) M_\odot$ unstable.
However, stars with mass above $\gtrsim$ 200 $M_\odot$ and very low metallicity (typical of population III) can avoid 
disruption and form a population of intermediate mass BHs with masses 
above $\sim 135 M_\odot$~\cite{rakavy,Barkat:1967zz,fraley,Heger:2001cd,Woosley:2007qp,Belczynski:2016jno,Woosley:2016hmi,Renzo:2020rzx,
Farmer:2019jed,Stevenson:2019rcw}.
It is not excluded that BHs within the mass gap might have an 
astrophysical origin due to hierarchical coalescences of smaller BHs 
\cite{Fishbach:2017dwv,Gerosa:2019zmo,Rodriguez:2019huv,Baibhav:2020xdf,Kimball:2020opk,Samsing:2020qqd,Mapelli:2020xeq}
, via direct collapse of a stellar merger between an evolved star and a main sequence companion~\cite{DiCarlo:2019fcq, 
DiCarlo:2020lfa}, by rapid mass accretion in primordial dense clusters \cite{rma}, or by beyond Standard Model 
physics that reduces the pair instability~\cite{Sakstein:2020axg}.

To understand if 
\name\ may have a primordial origin, one has to assess whether the corresponding merger rate is in 
agreement  with  observations and with the corresponding value of $f_\PBH$ allowed by the current experimental 
constraints~\cite{Carr:2020gox}.  
 We will perform our analysis for two scenarios:
{\it i)} the PBH mass function is determined only by the event \name;
{\it ii)} the PBH mass function is determined along the lines of Ref.~\cite{paper3}, i.e. assuming all the detections
by LIGO/Virgo (including \name) are originated by PBH mergers.

\vskip 0.3cm
\noindent
\paragraph{PBH Mass function.}
\noindent
Several models have been proposed to describe how PBHs form. One of the most likely scenarios 
is 
based on the collapse of sizeable overdensities in the radiation dominated epoch~\cite{s00,s0,s1,s2,s3}. The 
characteristic properties of the density perturbations (arising from curvature perturbations generated during 
inflation) are imprinted in the PBH mass distribution, whose shape, at formation redshift $z_\ii$, is often parametrised 
by a lognormal function
\begin{equation}
\label{psi}
\psi (M,z_\ii) = \frac{1}{\sqrt{2 \pi} \sigma M} {\rm exp} \left(-\frac{{\rm log}^2(M/M_*)}{2 \sigma^2} 
\right)\,,
\end{equation}
in terms of its central mass $M_*$ and width $\sigma$, arising from a symmetric peak 
in the curvature power spectrum~\cite{mf1,mf2}. 

\begin{figure}[t!]
	\includegraphics[width=\columnwidth]{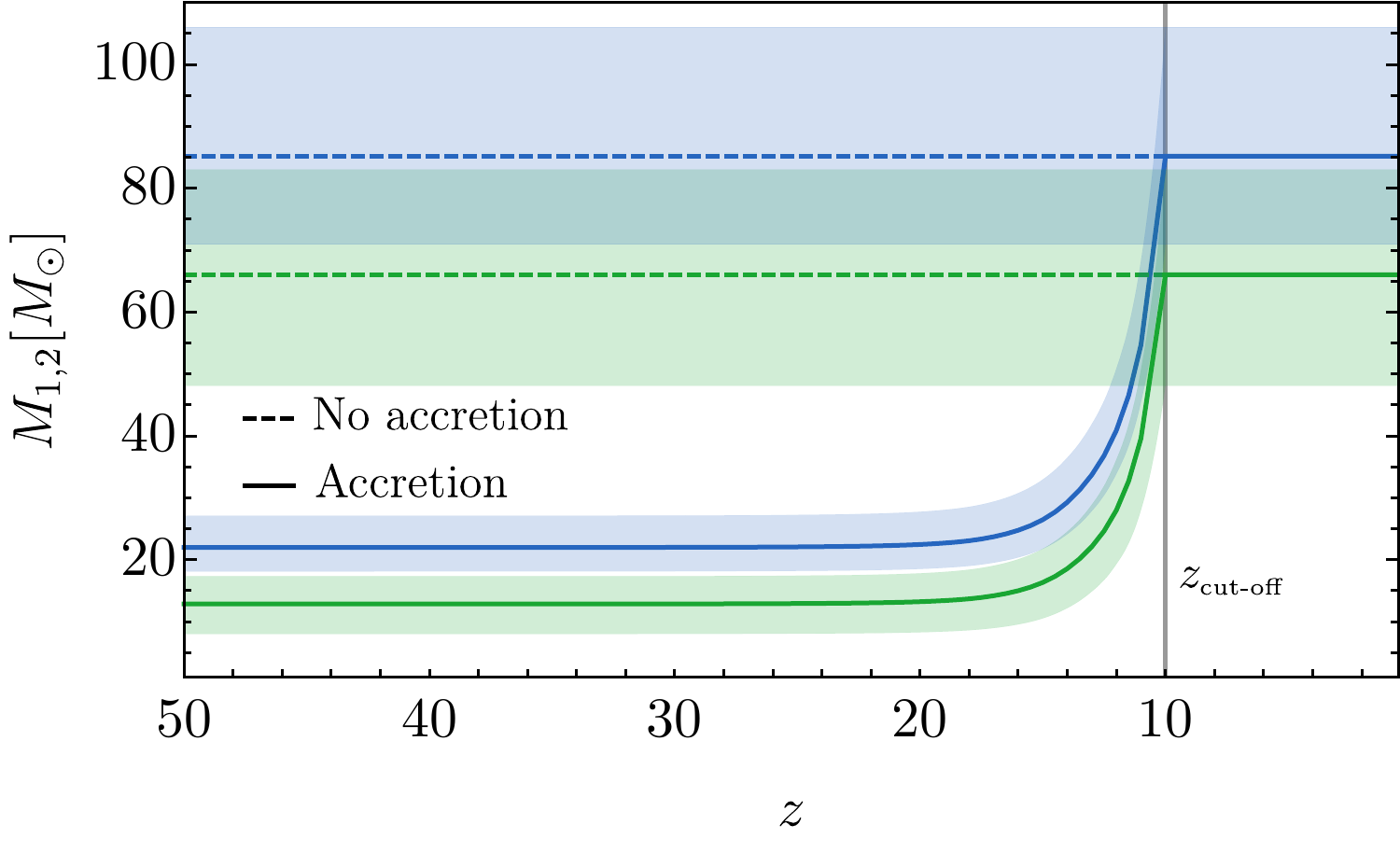}
	\caption{Evolution of the PBH masses for \name\ without and with accretion ($z_\text{\tiny 
cut-off}=10$).Green (blue) bands correspond to the primary (secondary) PBH mass within $90\%$ CL.}
	\label{fig1}
\end{figure}

If accretion is inefficient, the mass function in Eq.~\eqref{psi} 
describes the PBH population at the time of merger. In the opposite case, however,  such mass 
function must be properly evolved~\cite{Ricotti:2007jk,Ricotti:2007au,zhang,paper1}. 
Indeed, PBHs in binaries may be affected by periods of baryonic mass accretion for 
masses larger than $\mathcal{O}(10) M_\odot$, in which the whole binary system is responsible for gas infall provided 
its corresponding Bondi radius is bigger than the typical size of the binary~\cite{paper1, paper3}. This implies 
that both PBHs experience gas accretion with individual mass variations as~\cite{paper1, paper3} 
\begin{align}
\dot M_1 = \dot M_\text{\tiny bin}  \frac{1}{\sqrt{2 (1+q)}}, \qquad \dot M_2 = \dot M_\text{\tiny bin}  \frac{\sqrt{q} }{\sqrt{2 (1+q)}},
\label{M1M2dotFIN}
\end{align}
depending on the mass ratio $q \equiv M_2/M_1 \leq 1$ and  the binary's Bondi-Hoyle mass accretion rate 
\be
 \label{R1bin}
\dot M_\text{\tiny bin} = 4 \pi \lambda m_H n_\text{\tiny gas} v^{-3}_\text{\tiny eff} M^2_\text{\tiny tot},
\ee
in terms of the binary effective velocity $v_\text{\tiny eff}$ relative to the baryons with cosmic mean density 
$n_\text{\tiny gas}$ and the hydrogen mass $m_H$. 
The accretion parameter $\lambda$ keeps into account the effects of 
gas viscosity,  Hubble expansion, and the coupling of the CMB radiation to the gas through Compton 
scattering~\cite{Ricotti:2007jk}. 
Current limits on the fraction of the DM comprised by PBHs, with masses larger than $\mathcal{O}(M_\odot)$, indicate 
that an additional DM halo is expected to form around the compact systems, acting as a catalyst that enhances the gas 
accretion rate~\cite{Ricotti:2007au,Adamek:2019gns,Mack:2006gz}. This effect is included in the parameter 
$\lambda$~\cite{paper1}. We account for the sharp decrease in the accretion efficiency around the epoch of structure 
formation~\cite{Hasinger:2020ptw,raidalsm,Ali-Haimoud:2017rtz} and for other uncertainties in the model (such as 
X-Ray pre-heating~\cite{Oh:2003pm}, details of the structure formation and feedbacks of local, global 
\cite{Ricotti:2007au,Ali-Haimoud:2016mbv} and mechanical type~\cite{Bosch-Ramon:2020pcz})
by setting a cut-off redshift 
$z_\text{\tiny  cut-off}= 10$ after which we neglect accretion. Based on previous detailed 
investigation~\cite{paper1,paper3}, we consider $z_\text{\tiny  cut-off}= 10$ as a benchmark case to study the impact 
of accretion on the PBH scenario. Our conclusions are not qualitatively affected by this choice.
In Fig.~\ref{fig1} we plot the evolution of the PBH masses for \name\ with and without accretion.

The presence of accretion induces several effects in the PBH population. First, it shifts
the mass distribution to higher masses and enhances its high-mass tail~\cite{paper1}. Second, it modifies in a redshift-dependent fashion the overall fraction 
of PBHs in DM, scaling like $f_\PBH (z)/f_\PBH(z_\ii) \sim \langle M(z) \rangle/\langle M (z_\ii)\rangle$ for 
sufficiently small initial abundances~\cite{paper2}.
Third, the PBH mass evolution generates a mass ratio distribution which is skewed toward 
$q\lesssim1$ with a width depending upon the initial PBH mass function~\cite{paper3}.
Finally, the infalling gas may carry angular momentum which strongly affects the geometry of the accretion flow and the 
PBH spins~\cite{bv}. 
Efficient angular momentum transfer is present when an accretion disk forms~\cite{Ricotti:2007au,Shakura:1972te, 
  NovikovThorne},  leading to a spin growth depending  on the mass accretion as
$\dot \chi_j = g (\chi_j) \dot M_j/ M_j$, in terms 
of a function $g(\chi_j)$ of the dimensionless spin parameter $\chi_j = 
J_j/M_j^2$ of the $j$-th BH~\cite{Bardeen:1972fi,Brito:2014wla,volo,paper3}, up to the limit dictated by radiation 
effects, $\chi_\text{\tiny max} = 
0.998$~\cite{thorne,Gammie:2003qi}. If accretion is not effective, the PBH would retain its initial spin which, in the 
standard formation scenario in radiation domination, is around the percent 
level or smaller~\cite{DeLuca:2019buf,Mirbabayi:2019uph}.

\begin{figure*}[t!]
	\includegraphics[width=0.87\columnwidth]{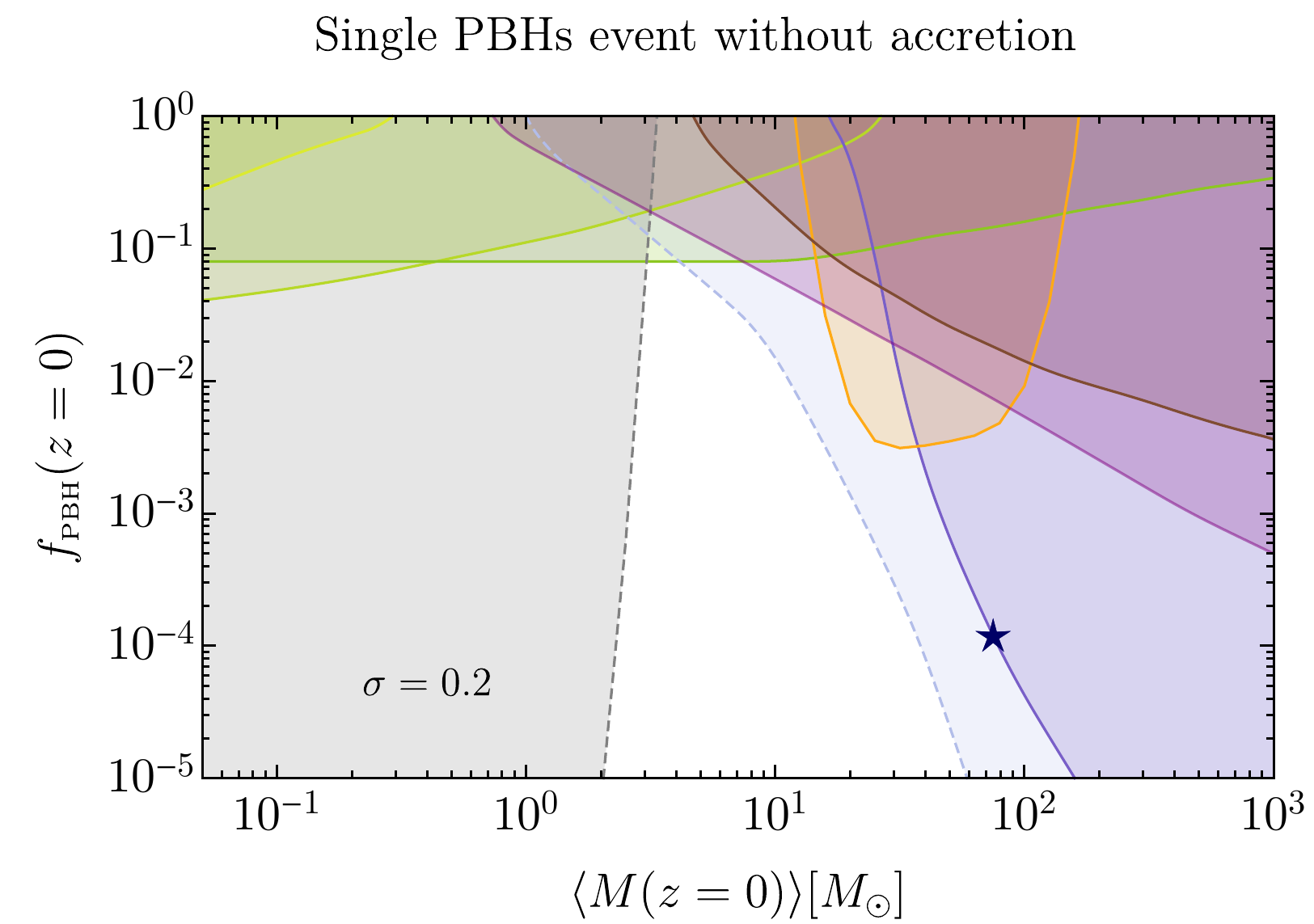}
	\includegraphics[width=1.1  \columnwidth]{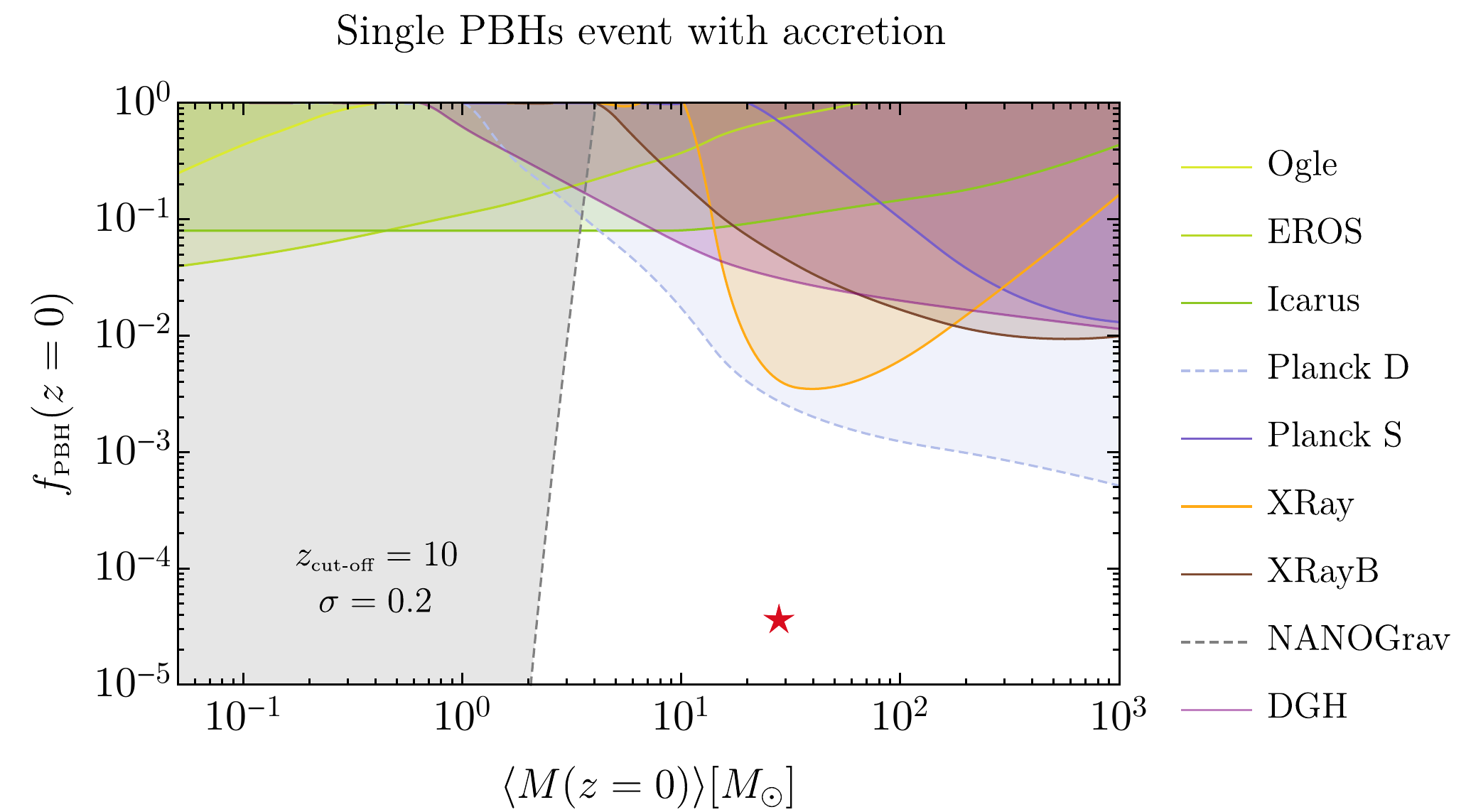}
	\caption{Constraints on $f_\PBH$ in the first scenario. The strongest ones come from the 
NANOGrav experiment~\cite{Chen:2019xse} (applicable only for PBHs formed from collapse of density 
perturbations and in the absence of non-Gaussianities, see~\cite{Nakama:2016gzw,Cai:2019elf}), CMB distortions 
(Planck~D and Planck~S)~\cite{Ali-Haimoud:2016mbv, serpico}, X-rays~(XRay)~\cite{Gaggero:2016dpq,Manshanden:2018tze}, 
X-ray binaries~(XRayB)~\cite{Inoue:2017csr},
EROS~\cite{Allsman:2000kg}, Icarus~\cite{Oguri:2017ock},  radio observations~(Ogle)~\cite{Niikura:2019kqi}, and Dwarf 
Galaxy Heating~(DGH)~\cite{kus}. Constraints on monochromatic PBH mass functions are adapted to extended mass functions 
with the techniques described in~\cite{mf2,Bellomo:2017zsr}. 
{\it Left:} The star indicates the PBH abundance needed to explain the single event \name\ in the PBH scenario 
without accretion. 
{\it Right:} Same as in the left panel but in the scenario with accretion, accounting for the change of PBH masses 
and abundances as discussed in detail in Ref.~\cite{paper2}. In particular, early-universe constraints 
NANOGrav and Planck (D/S) are computed with the high-redshift mass function and mapped to low redshift, 
whereas late-time universe constraints are evaluated with the mass function evolved up to low redshift.
}
	\label{singleeventconstraint}
\end{figure*}

\vskip 0.3cm
\noindent
\paragraph{PBH merger rate.}
\noindent
The PBH abundance and mass function at the formation epoch determine the probability that a two-body system decouples from the Hubble flow in the early universe and 
forms a binary. Both the surrounding population of PBHs and the density perturbations generate a torque on the binary 
system, which impacts the initial distribution of the semi-major axis and eccentricity. 
The PBH differential merger rate is~\cite{raidal}
\begin{align}
\label{mergerrate}
&\d R_{\text{\tiny  no-acc}} = \frac{1.6 \cdot 10^6}{{\rm Gpc^3 \, yr}} f_\PBH^{\frac{53}{37}} (z_\ii) 
\eta_\ii^{-\frac{34}{37}}   \lp \frac{t}{t_0} 
\rp^{-\frac{34}{37}}  
\lp \frac{M^\ii_\text{\tiny tot}}{M_\odot} \rp^{-\frac{32}{37}}  \nonumber \\
& \times S\lp M^\ii_\text{\tiny tot}, f_\PBH (z_\ii)  \rp
\psi(M^\ii_1, z_\ii) \psi (M^\ii_2, z_\ii)\d M^\ii_1 \d M^\ii_2,
\end{align}
where $\eta_\ii = \mu^\ii/M^\ii_\text{\tiny tot}$, $\mu^\ii = M^\ii_1 M^\ii_2/M^\ii_\text{\tiny tot}$, 
$M^\ii_\text{\tiny tot} = M^\ii_1 + M^\ii_2$, and $t_0$ is the current age of the universe.
The suppression factor $S$ accounts for the possible binary disruption by the surrounding PBHs and matter density 
perturbations~\cite{raidal}.
While the merger rate is not affected by fly-bys in the late-time universe~\cite{Young:2020scc}, there could be an 
additional suppression due to the disruption of binaries caused by other PBHs 
forming early sub-structures, which is effective only if $f_\PBH\gsim 0.1$~\cite{raidal,ver,Jedamzik:2020ypm,j2}.
However, the value of $f_\PBH$  necessary to explain \name\  is at 
least two orders of magnitude smaller. Notice also that, for such small abundances, PBH clustering 
is irrelevant~\cite{DeLuca:2020jug};
similarly, according to the results of $N$-body simulations~\cite{inman}, for $f_\PBH \lesssim z 
\cdot 10^{-4}$ PBHs are not clustered and therefore the stringent bounds from CMB distortions~\cite{Ali-Haimoud:2016mbv, 
serpico}, whose relevant physics occurs at $z = (300 \div 600)$, may not be evaded by clustering arguments. 
Finally, in the LIGO/Virgo band the fraction of PBHs generated by previous mergers is 
$\sim\mathcal{O}(10^{-2} f^{16/37}_\PBH)$ and second-generation mergers are 
negligible~\cite{paper1,Liu:2019rnx,Wu:2020drm}.

Accretion enhances the merger rate as~\cite{paper3}
\begin{align}
\label{mergerrateacc}
 \d R_\text{\tiny  acc} 
&= \d R_{\text{\tiny  no-acc}}  \lp \frac{ M_\text{\tiny tot}(z_\text{\tiny cut-off})}{M_\text{\tiny tot} (z_\text{\tiny i}) } \rp^{\frac{9}{37}} 
\lp\frac{\eta (z_\text{\tiny cut-off})}{\eta (z_\text{\tiny i}) }  \rp^{\frac{3}{37}}
\nonumber \\
& \times  
  \exp\llp \frac{12}{37} \int_{t_\ii} ^{t_\text{\tiny cut-off}} \d t \lp \frac{\dot M_\text{\tiny tot}}{M_\text{\tiny tot}} + 2 \frac{\dot \mu}{\mu} \rp \rrp.
\end{align}
The corrective factors account  for the hardening of the binary at $z<z_\text{\tiny 
cut-off}$~\cite{paper3,Caputo:2020irr} and the augmented energy loss through GW emission 
\cite{Peters:1963ux,Peters:1964zz}, both due to the mass accretion of each PBHs.

An additional contribution comes from binaries formed in the late-time universe, whose
rate density is~\cite{Bird:2016dcv,Ali-Haimoud:2017rtz} 
\begin{equation}
{\cal V}_\text{\tiny LU}\simeq 10^{-8} \left(\frac{f_\PBH}{10^{-3}}\right)^{53/21}  {\rm Gpc}^{-3}\,{\rm yr}^{-1}.
\end{equation}
This channel always gives a subdominant contribution with respect to the early-universe one.  Given the small 
abundances, PBH clustering does not increase the late-time universe merger rate~\cite{DeLuca:2020jug}, which we 
therefore neglect.

The rate of events reads~\cite{domi,Finn:1992xs, Berti:2004bd}
\be
\mathcal{R} = \int\d z \d M_1 \d M_2  \frac{p_\text{\tiny det}(z,M_1,M_2)}{1+z} \frac{\d V_c(z)}{\d z} \frac{\d 
R_\text{\tiny (no-)acc}}{\d M_1 \d M_2}\,,
\ee
where $V_c(z)$ indicates the comoving volume per unit redshift. 
The factor $p_\text{\tiny det}(z,M_1,M_2)$ accounts for 
the detection probability of a binary averaged over the source orientation and as a function of the 
signal-to-noise ratio~(SNR). We adopt the detectability threshold ${\rm SNR}=8$, the O3 noise power 
spectral density~\cite{domi,psd1,psd2,NN1,NN2}, and the IMRPhenomD
waveform model~\cite{IMRPhenomD}.
 The latter neglects precession, higher harmonics, and eccentricity, which could be relevant for GW190521.
Using the IMRPhenomXPHM waveform model~\cite{Pratten:2020ceb}, we checked that precession and higher harmonics affect 
the SNR computation at percent level, which is much smaller than the waveform modelling and prior systematics for this 
event, leading most notably to uncertainties on the binary masses, spins, and 
eccentricity~\cite{neweventsLonger,Romero-Shaw:2020thy,Gayathri:2020coq,Nitz:2020mga}.

\begin{figure*}[t!]
	\includegraphics[width=0.92 \columnwidth]{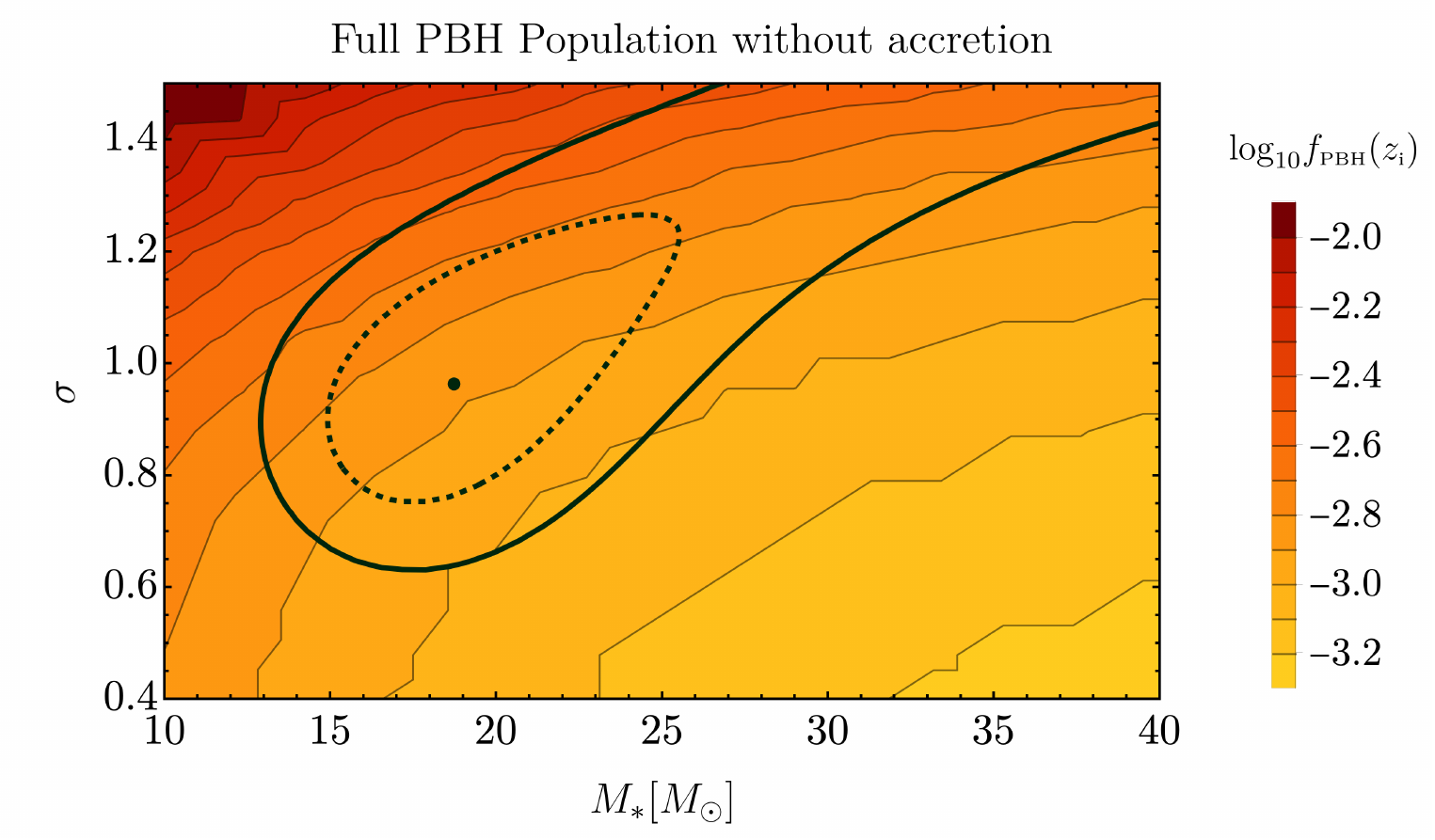}
	\includegraphics[width=0.95 \columnwidth]{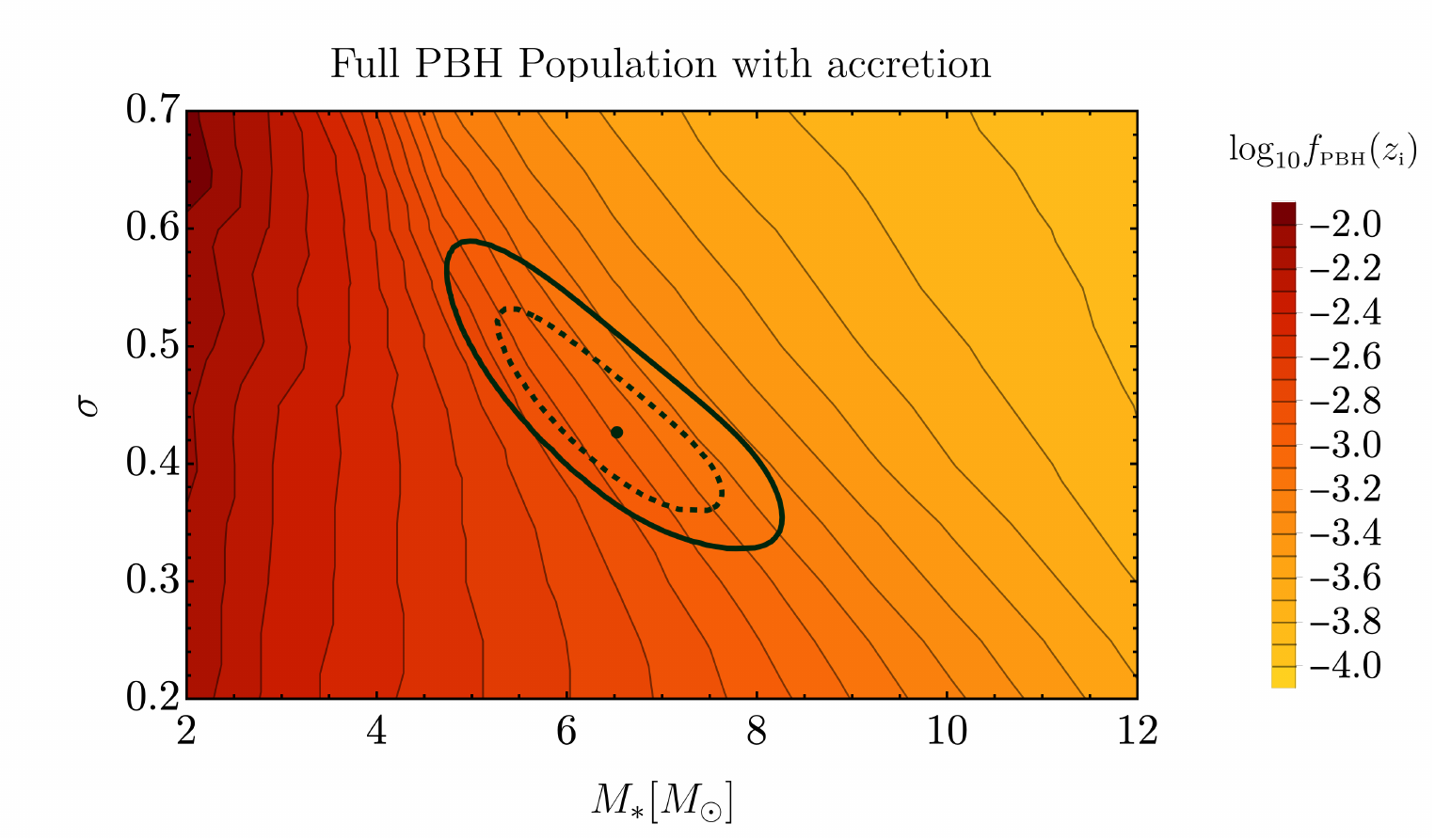}
	\caption{Likelihood in the PBH mass-function parameter space in the second scenario.
	Dashed (Solid) lines correspond to $1\sigma\ (2 \sigma)$ contours.
	{\it Left:}~No accretion scenario.
	{\it Right:}~Accretion scenario with $z_\text{\tiny cut-off}=10$.}
	\label{fullpoplikelihood}
\end{figure*}

\begin{figure*}[t!]
	\includegraphics[width=0.888 \columnwidth]{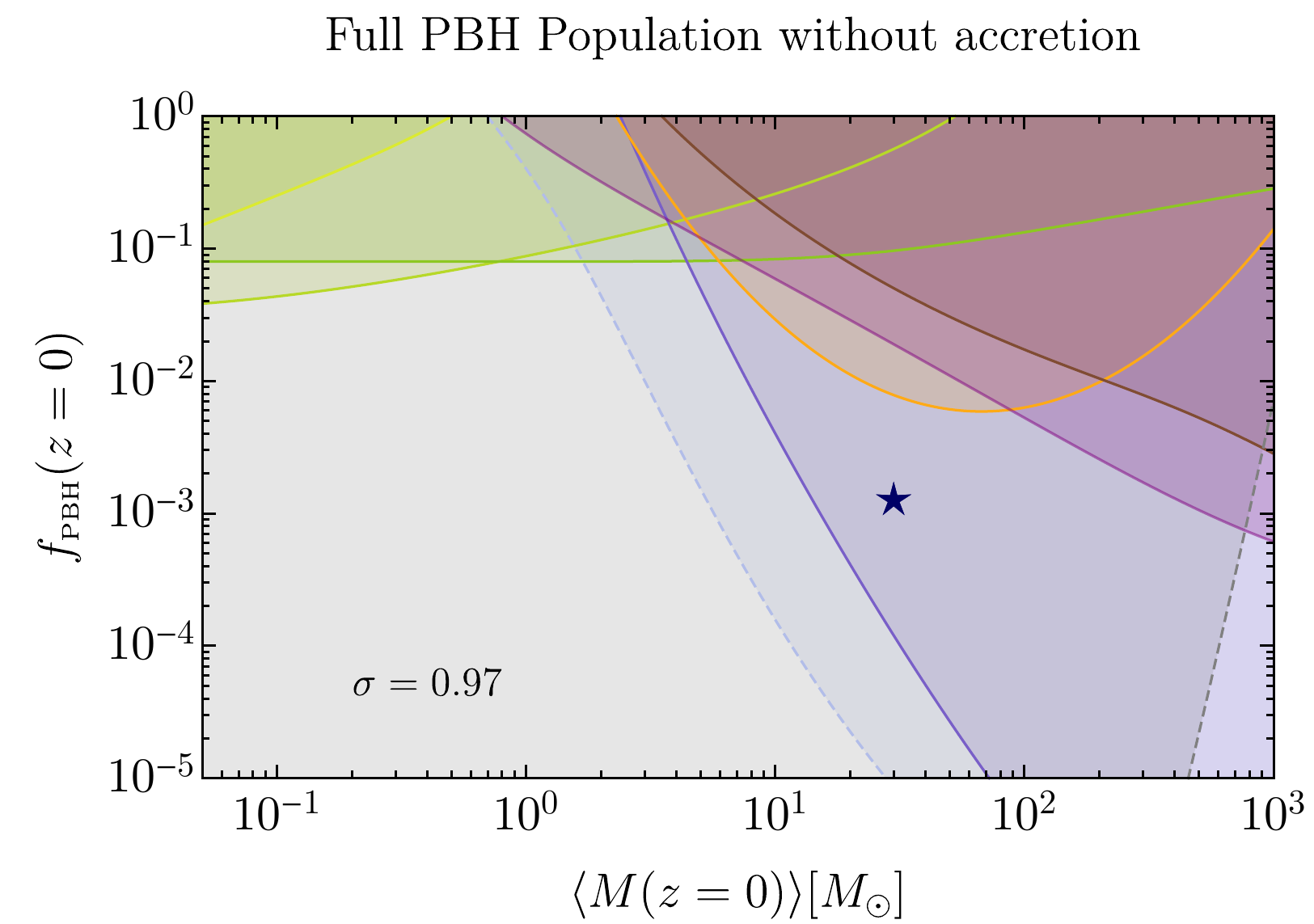}
	\includegraphics[width=1.12 \columnwidth]{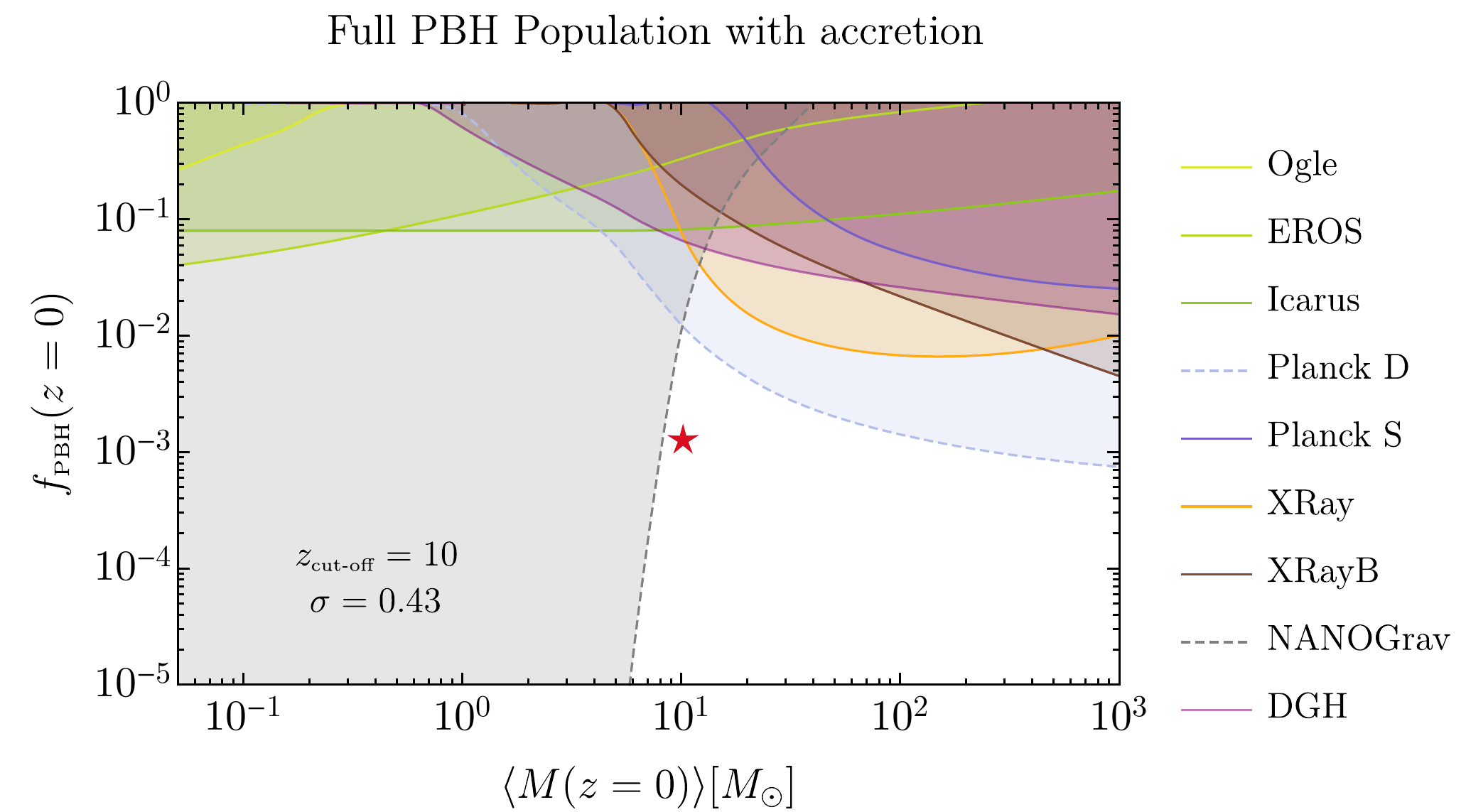}
	\caption{Same as in Fig.~\ref{singleeventconstraint} for the second scenario. The stars indicate $f_\text{\tiny PBH}$ needed to 
explain \name.
	Notice that the NANOGrav constraint, relevant only below $M_\odot$ for a monochromatic 
population~\cite{Chen:2019xse}, is able to constrain wide enough mass functions peaked at much larger masses.
}
	\label{fullpopconstraint}
\end{figure*}

\vskip 0.3cm
\noindent
\paragraph{\name\ as a single PBH event without accretion.} In this scenario the mass function is only determined by the 
considered event. We
choose the representative values $M_*=75 M_\odot$ and $\sigma=0.2$; different choices would result in a change of the 
merger rates only by ${\cal O}(1)$ factor. 
As the rate scales indicatively as $\mathcal{R} \propto f_\PBH^{53/37}$, the abundance is found to be relatively stable 
with respect to reasonable changes of the mass function.
From Eq.~\eqref{mergerrate}, we estimate the abundance required to match the observed event rate 
which, given the current high uncertainties~\cite{newevents,neweventsLonger}, we assume to be 
$\mathcal{R} \simeq 1 / {\rm yr}$. This yields $f_\PBH \simeq 1.2 \cdot 10^{-4}$. In Fig.~\ref{singleeventconstraint} 
we compare this value with the constraints, showing a tension with the bounds coming from CMB distortions.

\vskip 0.3cm
\noindent
\paragraph{\name\ as a single PBH event with accretion.}
In this scenario the two PBHs evolve from initial masses of $M_1^\text{\tiny i} \simeq 22 M_\odot$ and 
$M_2^\text{\tiny i} \simeq 13 M_\odot$, respectively (Fig.~\ref{fig1}).
We choose a representative narrow initial mass function with $M_*=18.5 M_\odot$ and $\sigma=0.2$. 
The abundance corresponding to the observed  merger rate is $f_\PBH (z_\ii)\simeq 2.5 \cdot 10^{-5}$
 ($f_\PBH (z=0)\simeq 3.7 \cdot 10^{-5}$), which is
 smaller than the one found in the previous section since accretion generally leads to an enhancement 
of the merger rate, see Eq.~\eqref{mergerrateacc}.
As shown in Fig.~\ref{singleeventconstraint}, this value of $f_\PBH$ is allowed by the current constraints, which are strongly relaxed in the presence of accretion~\cite{paper2}.

\noindent

\vskip 0.3cm
\noindent
\paragraph{\name \ as an event within the full PBH population.}  Finally, we perform a 
population-driven analysis assuming that all observations (O1/O2, GW190412, GW190815, \name) have a 
primordial origin.  Following the procedure in Ref.~\cite{paper3}, we implement a maximum-likelihood 
analysis  to find the PBH mass function at formation, see Fig.~\ref{fullpoplikelihood}.

In the case without relevant accretion, the best-fit values are $M_* = 19 M_\odot$ and $\sigma = 0.97 $.
The presence of the additional event \name \ with high masses causes a preference for a broader mass function.
We do not observe, however, a dramatic change in the best-fit parameters relative
to those found in Ref.~\cite{paper3}, since \name \ is  only one newly added event out of 13 
observations released to date, and not particularly loud (${\rm SNR}\approx14.7$)~\cite{newevents}. 
One can notice a characteristic trend for $f_\PBH$. Lighter PBHs reduce the fraction of detectable 
binaries. Also, for wider mass functions, the merger rate of binaries with $M_\text{\tiny 
tot}\gg M_*$ is suppressed~\cite{raidal}. Both effects require a larger $f_\PBH$ to fit the data. 
The shape of the $1\sigma$/$2\sigma$ contours can be explained as follows. As $M_*$ moves away from its best fit, 
a wider mass function is needed. 
A lower  $M_*$ tail is absent due to the suppression factor decreasing the merger rate for the observed masses 
$M_\text{\tiny tot} \gg M_*$, while a large $M_*-\sigma$ tail is still allowed since the suppression 
factor reduces the unwanted contribution of very large masses.

Fixing the merger rate to the observed one yields a value of $f_\PBH = 1.3 \cdot 10^{-3}$, which is already ruled out 
by the (conservative) bound coming from Planck S, and possibly by NANOGrav (see caption of Fig.~\ref{fullpopconstraint}). 
The observable rate of events with masses in the mass gap or at least as extreme as \name\  is
\begin{align}
&\mathcal{R} (M_{1,2}> 65 M_\odot) \simeq 7 \cdot 10^{-2} /{\rm yr},
\nonumber
\\
&\mathcal{R} (M_{1}> 85 M_\odot,M_{2}> 65 M_\odot) \simeq 2 \cdot 10^{-2}/{\rm yr}.
\end{align}
This proves \name\ to be an outlier of the population and not generally expected in the 
full population scenario without accretion.

In the case with sizeable accretion up to $z_\text{\tiny cut-off}=10$, the best-fit parameters are $M_* = 
6.5  M_\odot$ and $\sigma = 0.43 $, see Fig.~\ref{fullpoplikelihood}.
 The contours showing the $1\sigma/2\sigma$ C.I. are found to be significantly narrower relative to the no-accretion 
case. This is due to a stronger dependence of the binary masses and merger rates to $M_*$ and 
$\sigma$ as a result of the large impact of accretion onto binaries with total mass above ${\cal O} (10) M_\odot$. 
For large $M_*$ and $\sigma$, the merger rate is highly enhanced as in 
Eq.~\eqref{mergerrateacc}, leading to smaller values of $f_\text{\tiny PBH}$, at variance with the 
no-accretion case. For small values of $M_*$, accretion gradually becomes less relevant and 
the behavior of $f_\text{\tiny PBH}$ tends towards the one observed in the left panel of Fig.~\ref{fullpoplikelihood}.
Unlike the no-accretion case, a larger value of $M_*$ is correlated with a narrower mass function, since accretion 
broadens the initial mass distribution~\cite{paper2}.
 Also in this case, the initial mass function does not differ significantly relative to the one inferred 
without \name.
  
Fixing the merger rate with the LIGO/Virgo observations restrains the abundance to be 
$f_\PBH (z_\ii)\simeq 9.7 \cdot 10^{-4}$ ($f_\PBH (z=0)\simeq 1.0 \cdot 10^{-3}$),
 which is compatible with current constraints, see Fig.~\ref{fullpopconstraint}.
The merger rate of mass gap events (or as massive as \name) is 
\begin{align}
&\mathcal{R} (M_{1,2}> 65 M_\odot) \simeq 1.1 /{\rm yr},
\nonumber
\\
&\mathcal{R} (M_{1}> 85 M_\odot,M_{2}> 65 M_\odot) \simeq 0.8/{\rm yr},
\end{align} 
which are compatible with the observed rate.
Therefore, in the full population scenario with accretion, \name \ is not only perfectly 
allowed, but also generally expected.
This conclusion is solid against modelling systematics (e.g., changes of the accretion rate) and is also conservative. 
Indeed, in the more realistic hypothesis in which some GW events have an astrophysical origin (e.g. those whose 
parameters are slightly in 
tension with the PBH scenario, but see Ref.~\cite{Bhagwat:2020bzh} for the role of priors in alleviating this tension), 
removing them from the analysis would make our conclusion on the viability of \name\ as a primordial binary
even more robust.

\noindent
\vskip 0.3cm
\paragraph{Conclusions.} 
Motivated by the existence of a mass gap in the spectrum of stellar-origin BHs, we have 
conducted an in-depth analysis to understand if the recently observed event \name \ can be explained by the PBH 
hypothesis. 
We have explored two opposite scenarios:
the first assumes that this event is the only one of primordial origin 
among the LIGO/Virgo catalogues; the second supposes that all GW events observed so far are primordial
and \name \ is part of the inferred PBH population. 

We found that if accretion is negligible throughout the PBH cosmological evolution and within the second hypothesis, \name \ would be an outlier of the population with a corresponding merger rate orders of magnitude below the observed one.
Furthermore, in both scenarios  without accretion, the abundance required to match the merger rates is always largely 
in tension with the bounds coming from CMB distortions and, within the second hypothesis, with the one possibly coming 
from NANOGrav observations. 

We showed, however, that accretion relaxes the tension with the upper bounds on $f_\PBH$ in both scenarios, allowing 
for a primordial origin of \name. 
Furthermore, mergers similar to the newly discovered mass gap event \name \ and with a rate compatible with this 
observation are predicted by a PBH population inferred under the assumption that all events seen by LIGO/Virgo are 
primordial.

If  accretion is relevant for the PBHs, the mass ratio should be close to unity and the individual 
spins should be non vanishing, with the spin of the secondary component always bigger than the one of the 
primary~\cite{paper3}. Even though no firm conclusion can be drawn due to the 
small SNR and the large uncertainties coming from different waveform models~\cite{neweventsLonger}, this pattern seems 
to be in agreement with the parameters measured for \name.
There is evidence that the individual spins of the event are large and likely lie on the orbital plane. 
While sizeable spins are in tension with the non-accreting scenario, they are easily explained when accretion is efficient.
Since the Bondi radii of the individual PBHs in the binary are comparable to the characteristic orbital 
distance, the accretion flow's geometry is complex: the orientations of individual accretion disks are independent 
and randomly distributed irrespective of the direction of the orbital angular momentum. Thus, the inferred values of 
the binary components' spins are consistent with the accreting PBH scenario.

The accretion model used in this study relies on a Newtonian approximation. General-relativistic effects 
might significantly increase the accretion rate in the presence of a large density contrast in the accretion 
flow~\cite{Cruz-Osorio:2020dja}. It would be very interesting to extend numerical simulations~\cite{Cruz-Osorio:2020dja}
to the case under consideration, i.e. a BH binary moving in a (geometrically complex) density profile. We stress,  
however, that the accretion rate decreases by several orders of magnitude for PBH masses $\lesssim 10 
M_\odot$~\cite{Ricotti:2007au}, and therefore even an order-of-magnitude increase in $\dot M$ does not change the 
conclusions drawn from Figs.~\ref{singleeventconstraint} and \ref{fullpopconstraint}.

Our results give strong motivation to reconsider the parameter estimation of GW190521 incorporating the 
correct PBH-motivated prior distributions to infer the binary parameters in the PBH scenario~\cite{Bhagwat:2020bzh}, 
and to perform a Bayesian comparison between the PBH scenario and astrophysical ones that may explain spinning 
binaries in the mass gap, like hierarchical 
mergers~\cite{Fishbach:2017dwv,Gerosa:2019zmo,Rodriguez:2019huv,Baibhav:2020xdf,Kimball:2020opk,Mapelli:2020xeq}
or others~\cite{DiCarlo:2019fcq, DiCarlo:2020lfa}.

\noindent

\vskip 0.3cm
\noindent
\paragraph{Acknowledgments.}
\noindent
Computations were performed at University of Geneva on the Baobab cluster. 
We acknowledge use of the software package {\tt pycbc}~\cite{pycbc}.
V.DL., G.F. and 
A.R. are supported by the Swiss National Science Foundation 
(SNSF), project {\sl The Non-Gaussian Universe and Cosmological Symmetries}, project number: 200020-178787.
V.D. acknowledges support by the Israel Science Foundation (grant no. 1395/16).
P.P. acknowledges financial support provided under the European Union's H2020 ERC, Starting 
Grant agreement no.~DarkGRA--757480, and under the MIUR PRIN and FARE programmes (GW-NEXT, CUP:~B84I20000100001), and 
support from the Amaldi Research Center funded by the MIUR program `Dipartimento di 
Eccellenza" (CUP:~B81I18001170001).

\bigskip


\end{document}